\documentclass[onecolumn,useAMS,usenatbib]{mn2e}

\usepackage{graphicx}
\usepackage{subfigure}
\usepackage{xcolor}

\def\bc{\begin{center}}
\def\ec{\end{center}}
\def\be{\begin{eqnarray}}
\def\ee{\end{eqnarray}}

\title[Finslerian MOND vs. observations of Bullet Cluster]{Finslerian MOND vs. observations of Bullet Cluster 1E0657-558}
\author[X. Li, M.-H. Li, H.-N. Lin and Z. Chang]
        {Xin Li$^{1,2}$\thanks{E-mail:lixin@ihep.ac.cn;},
         Ming-Hua Li$^{1}$\thanks{E-mail:limh@ihep.ac.cn (corresponding author);},
         Hai-Nan Lin$^{1}$\thanks{E-mail: linhn@ihep.ac.cn;},
         and Zhe Chang$^{1,2}$\thanks{E-mail: changz@ihep.ac.cn}\\
$^{1}$Institute of High Energy Physics, Chinese Academy of Sciences, 100049 Beijing, China\\
$^{2}$Theoretical Physics Center for Science Facilities, Chinese Academy of Sciences, 100049 Beijing, China}
\begin{document}


\pagerange{\pageref{firstpage}--\pageref{lastpage}} \pubyear{2002}

\maketitle

\label{firstpage}

\begin{abstract}
It is known that theory of MOND with spherical symmetry cannot account for the convergence $\kappa$-map of Bullet Cluster 1E0657-558.
In this paper, we try to set up a Finslerian MOND, a generalization of MOND in Finsler spacetime. We use $Ric=0$ to obtain the gravitational vacuum field equation in a four-dimensional Finsler spacetime.
To leading order in the post-Newtonian approximation, we obtain the explicit form of the Finslerian line element. It is simply the Schwarzschild's metric except for the Finslerian rescaling coefficient $f(v)$ of the radial coordinate $r$, i.e. $R=f(v(r))r$.
By setting $f(v(r))=(1-\sqrt{a_0r^2/GM})^{-1}$, we obtain the famous MOND in a Finslerian framework. Taking a dipole and a quadrupole term into consideration, we give the convergence $\kappa$ in gravitational lensing astrophysics in our model. Numerical analysis shows that our prediction is to a certain extent in agreement with the observations of Bullet Cluster 1E0657-558. With the theoretical temperature $T$ taking the observed value 14.8 keV, the mass density profile of the main cluster obtained in our model is the same order as that given by the best-fit King $\beta$-model.

\end{abstract}


\section{Introduction}
It has long been noticed that according to Newton's inverse-square law of gravity, the observed baryonic matter cannot provide enough force to attract the matter of the galaxies \citep{Oort1932,Zwicky1933}. This inconsistency has been confirmed by a large number of observation in the past thirty years, to name a few, the velocity dispersions of dwarf Spheroidal galaxies \citep{Vogt} and the flat rotation curves of spiral galaxies \citep{Rubin,Walter}, {\it et al.}. Postulating that galaxies are surrounded by massive, non-luminous dark matter is the most widely adopted way to solve the problem \citep{de Blok}. The dark matter hypothesis has dominated astronomy and cosmology for almost 80 years. However, up to now, no direct observations have been firmly tested.

Some models have been built as an alternative of the dark matter hypothesis. Their main
ideas are to assume that the Newtonian gravity or Newton's dynamics
is invalid on galactic scales. The most successful and famous model is MOND \citep{Milgrom}. It assumes that the
Newtonian dynamics does not hold on galactic scales. The MOND paradigm is based on the following assumptions:~(i) It introduces a new physical constant $a_0=1.2\times10^{-8} {\rm cm/s^2}$. (ii) The law of gravity returns to Newton's gravity while $a_0\rightarrow0$. (iii) The law of gravity is given as $a_M=\sqrt{GMa_0}/r$ in the deep-MOND limit, $a_0\rightarrow\infty$.
As a phenomenological model, MOND explains well the flat rotation
curves of thousands of spiral galaxies with a simple formula and a universal constant. In particular, it naturally
gives the well-known global scaling relation for spiral
galaxies, the Tully-Fisher relation \citep{TF}. The Tully-Fisher relation is an empirical relation between the
total luminosity of a galaxy and the maximum rotational speed. It is of the form $L\propto v^a_{\rm max}$, where $a\approx4$, if the luminosity is measured in the near-infrared region. Tully and Pierce \citep{Tully} showed that the Tully-Fisher relation appears to be convergent in the near-infrared region. McGaugh \citep{McGaugh} investigated the Tully-Fisher relation for a large sample of galaxies, and concluded that the Tully-Fisher relation is a fundamental relation between the total baryonic mass and the rotational speed. MOND \citep{Milgrom} predicted that the rotational speed of galaxy has an asymptotic value $v^4_{~|r\rightarrow\infty}=GMa_0$, which explains the Tully-Fisher relation.

By introducing several scalar, vector and tensor fields, Bekenstein \citep{Bekenstein} rewrote the MOND into a covariant formulism (TeVeS). He showed that the MOND satisfies all four classical tests of Einstein's general relativity in Solar system. However, MOND still faces challenges. The strong and weak gravitational lensing observations of Bullet Cluster 1E0657-558 \citep{Bullet} cannot be explained by MOND and its Bekenstein's relativistic version \citep{Angus}. The ICM (intracluster medium) gas accounts for most of the Bullet Cluster's mass. Clowe {\it et al}. \citep{Bullet} had reconstructed the surface mass density $\Sigma(x,y)$ from the Chandra space satellite X-ray image of the ICM gas. Moffat {\it et al}. \citep{Moffat} had shown that the $\Sigma$-map of the ICM gas of the main cluster can be well fitted with a King $\beta$-model density profile. The King $\beta$-model is a radial distribution of the mass density for a nearly isothermal and isotropic gas sphere. On the other hand, Clowe {\it et al}. \citep{Bullet} had reconstructed the convergence $\kappa$-map from the strong and weak gravitational lensing survey. The $\kappa$-map indicates that additional gravitational force is needed for explaining the Bullet Cluster. The center of gravitational force deviates from the center of the ICM gas. And the distribution of gravitational force does not possess spherical symmetry. Most of theories of modified gravity, such as Bekenstein's relativistic version of MOND, only consider radial force. Also, most of the mass density profile of dark matter, such as the NFW profile \citep{Navarro}, only contrive radial (isotropic) distributions. All of them cannot explain the observations of the Bullet Cluster.

The distribution of gravitational force in Bullet Cluster is anisotropic. To describe anisotropic force, one should introduce the multipole fields. The dipole contribution vanishes if one takes the center of ICM gas as the coordinate origin. Monopole contribution plus quadrupole contribution are needed to account for the observations of Bullet Cluster. In fact, Milgrom gave a Quasi-linear formulation of MOND (QUMOND)\citep{QUMOND}, which involves the quadrupole contribution. Usually, MOND effects vanish in Newtonian regime. However, Milgrom showed that quadrupole effect appears even in high-acceleration systems. Besides, Angus \textit{et al.} presented an $N$-body code for solving the modified Poisson equation of QUMOND \citep{Angus2012}. They used it to compute rotation curves for a sample of five spiral galaxies from the THINGS sample \citep{Walter} and concluded that taking gas scale-heights of the gas-rich dwarf spiral galaxies (and stellar scale-heights of stellar dominated galaxies) as free parameters is vital to make precise conclusions about MOND. Other interesting results were also obtained in their work.

On the other hand, besides Bekenstein's TeVeS, there are other `MONDian' theories (for example, the Einstein-aether theory \citep{Zlosnik}). Both the Bekenstein's and the Einstein-aether theory admit a preferred reference frame and broken local Lorentz invariance. It can be reasonably inferred that the local Lorentz invariance violation (LIV) is an intrinsic feature of MOND. If this is  acknowledged, there follows a conclusion: the space structure near a galaxy is not Minkowskian even at long distances from the galaxy center. It depends on the rotational velocity of the galaxy considering the relationship between the Tully-Fisher relation and MOND.

Finsler gravity, which is based on Finsler geometry, has the features mentioned above. Thus it is natural to postulate that Finsler gravity is a covariant formulism of MOND. Finsler geometry \citep{Book by Bao} is a natural generalization of Riemannian geometry with the latter as a special case. The length element of an arc on a Finslerian manifold depends not only on position but also on velocity, which induces anisotropy. Preservation of the fundamental principles is a prerequisite for Finsler gravity as well as the results of general relativity. A new geometry (i.e. the Finsler geometry) involves new spacetime symmetries. Kostelecky \citep{Kostelecky} has shown that LIV is closely related to the Finslerian geometry. Effective field theories have been studied in his paper for explicit LIV effects in Finslerian spacetime.

In \citep{Finsler GW}, we presented the vacuum field equation in Finsler gravity, and have given the solution of field equation under weak field approximation. In \citep{Finsler MOND}, we presented the Newtonian limit in Finsler gravity and a covariant formulism of MOND. We studied the spacetime structure of MOND with properties of Tully-Fisher relation and Lorentz invariance violation. A Finsler spacetime has less symmetry than a Minkowski one \citep{Finsler PF}. Multipole effects such as dipole and quadrupole contributions, which embody space anisotropies, should be considered in Finsler gravity. In this paper, we try to construct a Finslerian MOND, a generalization of MOND in Finsler spacetime, and use it to explain the observations of Bullet Cluster.

The rest of paper are organized as follows. Section 2 is dedicated to the theory used for the numerical analysis and is separated into five parts: Section 2.1 is about the basic concepts of Finsler geometry; in Section 2.2, we discuss the null set for massless particles; in section 2.3, we extend Pirani's argument to a Finsler spacetime to obtain the gravitational vacuum field equation in Finsler gravity; in Section 2.4, the Newtonian limit in Finsler spacetime is presented; in Section 2.5, under post-Newtonian approximation we give the Finsler structure. In Section 3, we give the convergence $\kappa$ in our model. Section 4 is about the numerical analysis which contains two parts: in Section 4.1, we consider the dipole and quadrupole contributions to the Finslerian MOND in the calculation of the convergence $\kappa$ of the Bullet Cluster; in Section 4.2, we obtain the mass density of the main cluster given the observed value of its surface temperature and compare it to the best-fit King $\beta$-model. Numerical results are presented in 2D as well as in 3D figures. Conclusions and necessary discussions are presented in Section 5. Demonstrations of certain approximations in Section 3 is presented in the Appendix.

\section[]{Formulism of Finsler gravity}
\subsection{Basic Concepts}
Finsler geometry is based on the so called Finsler structure $F$. $F$ is a non-negative real function which has the property
$F(x,\lambda y)=\lambda F(x,y)$ for all $\lambda>0$, where $x$ represents position
and $y\equiv\frac{dx}{d\tau}$ represents velocity. The fundamental tensor is given as \citep{Book
by Bao}
 \begin{equation}
 g_{\mu\nu}\equiv\frac{\partial}{\partial
y^\mu}\frac{\partial}{\partial y^\nu}\left(\frac{1}{2}F^2\right).
\end{equation}
The arc length in Finsler space is given as
\begin{equation}
\label{integral length}
\int^r_sF(x^1,\cdots,x^n;\frac{dx^1}{d\tau},\cdots,\frac{dx^n}{d\tau})d\tau~.
\end{equation}
A more detailed discussion about $F$ can be found in Section 5.
Hereafter, we adopt the following index gymnastics: Greek indices in lower case run from 1 to 4, while Latin indices in lower case (except the alphabet $n$) run from 1 to 3.

The parallel transport
has been studied in the framework of Cartan
connection \citep{Matsumoto,Antonelli,Szabo}. The notation of parallel
transport on a Finsler manifold means that the length
$F\left(\frac{dx}{d\tau}\right)$ is constant.
The geodesic equation on a Finslerian manifold is given as \citep{Book by Bao}
\begin{equation}
\label{geodesic}
\frac{d^2x^\mu}{d\tau^2}+2G^\mu=0,
\end{equation}
where
\begin{equation}
\label{geodesic spray}
G^\mu=\frac{1}{4}g^{\mu\nu}\left(\frac{\partial^2 F^2}{\partial x^\lambda \partial y^\nu}y^\lambda-\frac{\partial F^2}{\partial x^\nu}\right)
\end{equation} are called the geodesic spray coefficients. $\tau$ is arc length on the Finsler manifold.
Obviously, if $F$ is a Riemannian metric, then
\begin{equation}
G^\mu=\frac{1}{2}\gamma^\mu_{~\nu\lambda}y^\nu y^\lambda,
\end{equation}
where $\gamma^\mu_{~\nu\lambda}$ is the Riemannian Christoffel symbol.
Since the geodesic equation (\ref{geodesic}) is directly
derived from the integral length
\begin{equation} L=\int
F\left(\frac{dx}{d\tau}\right)d\tau,
\end{equation} the inner product
$\left(\sqrt{g_{\mu\nu}\frac{dx^\mu}{d\tau}\frac{dx^\nu}{d\tau}}=F\left(\frac{dx}{d\tau}\right)\right)$
of two parallel transported vectors is preserved.

On a Finslerian manifold, there exists a linear connection~-~the Chern connection \citep{Chern}. It is of torsion freeness and almost metric-compatible,
\begin{equation}\label{Chern connection}
\Gamma^{\alpha}_{~\mu\nu}=\gamma^{\alpha}_{~\mu\nu}-g^{\alpha\lambda}\left(A_{\lambda\mu\beta}\frac{N^\beta_{~\nu}}{F}-A_{\mu\nu\beta}\frac{N^\beta_{~\lambda}}{F}+A_{\nu\lambda\beta}\frac{N^\beta_{~\mu}}{F}\right),
\end{equation}
where $\gamma^{\alpha}_{~\mu\nu}$ is the formal Christoffel symbols of the second kind with the same form of Riemannian connection. $N^\mu_{~\nu}$ is defined as
$N^\mu_{~\nu}\equiv\gamma^\mu_{~\nu\alpha}y^\alpha-A^\mu_{~\nu\lambda}\gamma^\lambda_{~\alpha\beta}y^\alpha y^\beta$
and $A_{\lambda\mu\nu}\equiv\frac{F}{4}\frac{\partial}{\partial y^\lambda}\frac{\partial}{\partial y^\mu}\frac{\partial}{\partial y^\nu}(F^2)$ is the Cartan tensor (regarded as a measurement of deviation from the Riemannian Manifold). In terms of the Chern connection, the curvature of Finsler space is given as
\begin{equation}\label{Finsler curvature}
R^{~\lambda}_{\kappa~\mu\nu}=\frac{\delta
\Gamma^\lambda_{~\kappa\nu}}{\delta x^\mu}-\frac{\delta
\Gamma^\lambda_{~\kappa\mu}}{\delta
x^\nu}+\Gamma^\lambda_{~\alpha\mu}\Gamma^\alpha_{~\kappa\nu}-\Gamma^\lambda_{~\alpha\nu}\Gamma^\alpha_{~\kappa\mu},
\end{equation}
where $\frac{\delta}{\delta x^\mu}=\frac{\partial}{\partial x^\mu}-N^\nu_{~\mu}\frac{\partial}{\partial y^\nu}$.

\subsection{The Null Set $F=0$ and Finslerian Special Relativity}
In Finsler geometry, the Finsler structure $F$ is defined as a non-negative $C^{\infty}$ function on the entire slit tangent bundle $TM\setminus0$, i.e. $F:TM\rightarrow [0,\infty )$. It ensures that the integral length (\ref{integral length}) always makes sense (since a negative arc length is not acceptable in mathematics). In physics, for a gravity theory, the quantity $F^2 d\tau^2$ represents the line element of spacetime (which is also called `proper time interval' in some references). A positive, zero and negative $F$ correspond to time-like, light-like (`null') and space-like curves respectively. For massless particles, the stipulation is $F=0$.

One should notice that many Finslerian geometric objects like Ricci scalar involves the Finsler structure $F$. It might be invalid to describe the massless particles at first glance. However, the ambiguities caused by $F = 0$ can be removed by re-parameterizing the formulae with some other parameter $\sigma$ such that $F(\sigma)\neq 0$. The property of Finsler structure $F(x,\lambda y)=\lambda F(x,y)$ guarantees that the length $L$ is independent of the choice of curve parameter. Under a given parameter change $\tau=C(\sigma)$, $\frac{d\sigma}{d\tau}>0$, the length $L$ is of the form
$L(\tau)=\int_s^r F\left(x,\frac{dx}{d\sigma}\frac{d\sigma}{d\tau}\right) d\tau=\int_s^r F\left(x,\frac{dx}{d\sigma}\right) d\sigma=L(\sigma)$, where $\tau$ and $\sigma$ are both curve parameters and $y\equiv dx/d\tau$ (or $y\equiv dx/d\sigma$). The same trick has been played in general relativity for massless particles which has $g_{\mu\nu} \frac{dx^\mu}{d\tau}\frac{dx^\nu}{d\tau}=0$ \citep{Weinberg}.

To construct Finslerian special relativity, one should study the symmetry of Finsler spacetime, i.e. the isometric group and Killing vectors. For projectively flat ($\alpha$,$\beta$) spacetime with constant flag curvature, this was done in \citep{Finsler PF}.

\subsection{Extension of Pirani's Arguments}
In this paper, we introduce the vacuum field equation in a way first discussed by Pirani \citep{Pirani, Rutz}. In Newton's theory of gravity, the equation of motion of a test particle is given as
\begin{equation}
\label{dynamic Newton}
\frac{d^2x^i}{d\tau^2}=-\eta^{ij}\frac{\partial \phi}{\partial x^i},
\end{equation}
where $\phi=\phi(x)$ is the gravitational potential and $\eta^{ij}=$ \textmd{diag}(+1,+1,+1) is the Euclidean metric. For an infinitesimal transformation $x^i\rightarrow x^i+\epsilon\xi^i$($|\epsilon|\ll1$), the equation (\ref{dynamic Newton}) becomes, to first order of $\epsilon$,
\begin{equation}
\label{dynamic Newton1}
\frac{d^2x^i}{d\tau^2}+\epsilon\frac{d^2\xi^i}{d\tau^2}=-\eta^{ij}\frac{\partial \phi}{\partial x^i}-\epsilon\eta^{ij}\xi^k\frac{\partial^2\phi}{\partial x^j\partial x^k}.
\end{equation}
Combining equations (\ref{dynamic Newton}) and (\ref{dynamic Newton1}), we obtain
\begin{equation}
\frac{d^2\xi^i}{d\tau^2}=\eta^{ij}\xi^k\frac{\partial^2\phi}{\partial x^j\partial x^k}\equiv\xi^kH^i_{~k}.
\end{equation}
For the vacuum field equation, one has $H^i_{~i}=\bigtriangledown^2\phi=0$.

In general relativity, the geodesic deviation gives a similar equation
\begin{equation}
\frac{D^2\xi^\mu}{D\tau^2}=\xi^\nu \tilde{R}^\mu_{~\nu},
\end{equation}
where $\tilde{R}^\mu_{~\nu}=\tilde{R}^{~\mu}_{\lambda~\nu\rho}\frac{dx^\lambda}{d\tau}\frac{dx^\rho}{d\tau}$. Here, $\tilde{R}^{~\mu}_{\lambda~\nu\rho}$ is the Riemannian curvature tensor. `$D$' denotes the covariant derivative along the curve $x^\mu(t)$. The vacuum field equation in general relativity gives $\tilde{R}^{~\lambda}_{\mu~\lambda\nu}=0$ \citep{Weinberg}. This implies that the tensor $\tilde{R}^\mu_{~\nu}$ is also traceless, $\tilde{R}\equiv\tilde{R}^\mu_{~\mu}=0$.

In Finsler spacetime, the geodesic deviation yields \citep{Book by Bao}
\begin{equation}
\frac{D^2\xi^\mu}{D\tau^2}=\xi^\nu R^\mu_{~\nu},
\end{equation}
where $R^\mu_{~\nu}=R^{~\mu}_{\lambda~\nu\rho}\frac{dx^\lambda}{d\tau}\frac{dx^\rho}{d\tau}$. Here, $R^{~\mu}_{\lambda~\nu\rho}$ is Finsler curvature tensor defined in (\ref{Finsler curvature}), `$D$' here denotes covariant derivative $\frac{D\xi^\mu}{D\tau}=\frac{d\xi^\mu}{d\tau}+\xi^\nu\frac{dx^\lambda}{d\tau}\Gamma^\mu_{~\nu\lambda}(x,\frac{dx}{d\tau})$. Since the vacuum field equations of Newton's gravity and general relativity are of similar forms, we may assume that vacuum field equation in Finsler spacetime has similar requirements as in the case of Netwon's gravity and general relativity. It implies that the tensor $R^\mu_{~\nu}$ in Finsler geodesic deviation equation should be traceless, $R^\mu_{~\mu}=0$. In fact, we have proved that the analogy of the geodesic deviation equation is valid at least in a Finsler spacetime of Berwald type \citep{Finsler DM}. We assume that this analogy still holds its validity in a general Finsler spacetime.

In Finsler geometry, there is a geometrical invariant --- the Ricci scalar $Ric$ . It is of the form \citep{Book by Bao}
\begin{equation}\label{predecessor flag curvature}
Ric\equiv R^\mu_{~\mu}=\frac{1}{F^2}\left(2\frac{\partial G^\mu}{\partial x^\mu}-y^\lambda\frac{\partial^2 G^\mu}{\partial x^\lambda\partial y^\mu}+2G^\lambda\frac{\partial^2 G^\mu}{\partial y^\lambda\partial y^\mu}-\frac{\partial G^\mu}{\partial y^\lambda}\frac{\partial G^\lambda}{\partial y^\mu}\right).
\end{equation}
The Ricci scalar depends only on the Finsler structure $F$ and is insensitive to the connection.
For a tangent plane $\Pi\subset T_xM$ and a non-zero vector $y\in T_xM$, the flag curvature is defined as
\begin{equation}
\label{flag curvature}
K(\Pi,y)\equiv\frac{g_{\lambda\mu}R^\mu_{~\nu}u^\nu u^\lambda}{F^2g_{\rho\theta}u^\rho u^\theta-(g_{\sigma\kappa}y^\sigma u^\kappa)^2},
\end{equation}
where $u\in\Pi$. The flag curvature is a geometrical invariant and a generalization of the sectional curvature in Riemannian geometry. The Ricci scalar $Ric$ is the trace of $R^\mu_{~\nu}$, which is the predecessor of flag curvature. Thus the value of Ricci scalar $Ric$ is invariant under the coordinate transformation.

Furthermore, the significance of the Ricci scalar $Ric$ is very clear. It plays an important role in the geodesic deviation equation \citep{Finsler GW,Finsler MOND,Book by Bao}. The vanishing of the Ricci scalar $Ric$ implies that the geodesic rays are parallel to each other. It means that it is vacuum outside the gravitational source.

Therefore, we have enough reasons to believe that the gravitational vacuum field equation in Finsler geometry has its essence in $Ric=0$. Pfeifer {\it et al}. \citep{Pfeifer1} have constructed gravitational dynamics for Finsler spacetime in terms of an action integral on the unit tangent bundle. The stipulation $Ric=0$ here is compatible with their results of gravitational field equation\footnote{The gravitational vacuum field equation given in  \citep{Pfeifer1} is $g^{F\,ab}\bar\partial_a\bar\partial_b\mathcal{R}-\frac{6}{F^2}\mathcal{R}+2g^{F\,ab}\big(\nabla_aS_b+S_aS_b+\bar\partial_a\nabla S_b\big)=0
$. The $S_a$-terms can be written as $S_a=\ell^{d}P_{d~ba}^{~b}$, where $\ell^d\equiv\frac{y^d}{F}$ and $P_{d~ba}^{~b}$ are the coefficients of the cross basis $dx \wedge \frac{\delta y}{F}$ \citep{Book by Bao}. Considering that $\mathcal{R}=R^a_{~ab}y^b=-R^a_{~dab}y^d y^b=F^2(\ell^d R^{~a}_{d~ab}\ell^b)=F^2(g^{ab}R_{ab})=F^2 Ric$ and dropping the $S_a$-terms (see the discussions about the `torsion' terms in next section), one can see that $Ric=0$ is one of the solutions of the above equation.}.

\subsection{The Newtonian Limit in Finsler Spacetime}
It is well known that the Minkowski spacetime is a trivial solution of the Einstein's vacuum field equation. In Finsler spacetime, the trivial solution of the vacuum field equation is called `locally Minkowski spacetime'. A Finsler spacetime is called a locally Minkowski spacetime if there is a local coordinate system $\{x^\mu\}$, with induced tangent space coordinates $\{y^\mu\}$, such that $F$ depends not on $x$ but only on $y$. The locally Minkowski spacetime is a flat spacetime in Finsler geometry. Using the formula (\ref{predecessor flag curvature}), one can see that a locally Minkowski spacetime is a solution of Finslerian vacuum field equation.

In  \citep{Finsler GW,Finsler MOND}, we assumed that the metric is close to the a locally Minkowski one $\eta_{\mu\nu}(y)$,
\begin{equation}\label{expand metric}
g_{\mu\nu}=\eta_{\mu\nu}(y)+h_{\mu\nu}(x,y),~~~|h_{\mu\nu}|\ll1\ ,
\end{equation}
considering that the gravitational field $h_{\mu\nu}$ is stationary (thus all time derivatives of $h_{\mu\nu}$ vanishes) and the particle is moving very slowly (i.e. $GM/r\ll 1$). The lowering and raising of indices are carried out by $\eta_{\mu\nu}$ and its matrix inverse $\eta^{\mu\nu}$.
We found from $Ric=0$, to first order of $h_{\mu\nu}$, that
\begin{equation}\label{static field eq}
\eta^{ij}\frac{\partial^2 h_{\alpha\beta}}{\partial x^i \partial x^j}y^\alpha y^\beta +\mathcal{O}\left(h_{\mu\nu} \right)=0.
\end{equation}

In general relativity, one uses post-Newtonian approximation to study the motion of particle \citep{Weinberg}. Before studying the motion of particle in Finsler gravity, we must deal with the concept of energy-momentum tensor in Finsler spacetime. It is well known that the energy-momentum tensor is conserved (in the sense of covariant differentiation) and symmetric in general relativity. However, this is not the case in Finsler gravity. The energy-momentum tensor is symmetric if the angular momentum is conserved \citep{GTM 93}. Generally, the symmetry of angular momentum is broken in Finsler spacetime \citep{Finsler PF}. Thus, the energy-momentum tensor is not symmetric in Finsler gravity. Similar situations appear in torsion gravity \citep{Hammond} in Riemann-Cartan geometry. Also, to satisfy the conservation law, besides the Ricci scalar, additional terms that represent the ``torsion effect" are needed in the field equation \citep{Pfeifer1}. Although these ``torsion" terms would cause a difficulty to understand Finsler gravity, they could fortunately be omitted. The reason is that these ``torsion" terms do not contribute to the geodesic deviation equation, which determines the motion of particles in Finsler geometry. Furthermore, we concentrate only on the motion of particle with zero spin in a weak gravitational field. Therefore, with similar steps to deduce the equation (\ref{static field eq}) in  \citep{Finsler GW,Finsler MOND}, and by making use of the post-Newtonian approximation, we obtain the gravitational field equation in Finsler gravity \footnote{The derivations in the rest of this and the next subsections, if not specifically pointed out, are accurate to the first-order of $h_{\mu\nu}$.}
\begin{eqnarray}\label{static field eq1}
\eta^{ij}\frac{\partial^2 h_{\alpha\beta}}{\partial x^i \partial x^j} +\mathcal{O}\left(h_{\mu\nu} \right)=-\kappa\left(T_{\alpha\beta}-\frac{1}{2}\eta_{\alpha\beta}T^\lambda_{~\lambda}\right).
\end{eqnarray}
$h_{00},h_{nn}$ are terms of the same order as $GM/r$, and the corresponding component of the energy-momentum tensor is $T_{00}$. Finsler gravity should reduce to general relativity, if the Finsler metric $g_{\mu\nu}$ reduces to a Riemannian one. Thus, we find from (\ref{static field eq1}) that
\begin{eqnarray}\label{static field eq2}
\eta^{ij}\frac{\partial^2 h_{00}}{\partial x^i \partial x^j}&=&-8\pi_F G\rho\eta_{00},\\
\label{static field eq3}
\eta^{ij}\frac{\partial^2 h_{nn}}{\partial x^i \partial x^j}&=&8\pi_F G\rho\eta_{nn},
\end{eqnarray}
where $\rho= T_{00}/\eta_{00}$ is the energy density of the gravitational source. In Finsler spacetime, the space volume of $\eta_{\mu\nu}(y)$ \citep{Book by Bao} is different from the one in Euclidean space. We used $\pi_F$ in (\ref{static field eq2},\ref{static field eq3}) to represent the difference, where
\begin{equation}
\pi_F\equiv\frac{3}{4}\int_{R=1}\sqrt{g}dx^1\wedge dx^2\wedge dx^3.
\end{equation}
$g\equiv det(\eta_{ij})$ is the determinant of $\eta_{ij}$. `$\wedge$' denotes the `wedge product'\footnote{In Subsection 2.2, we have discussed the local symmetry of Finsler spacetime. It manifests that the symmetry of locally Minkowski spacetime is different from the Minkowski spacetime. The space length that determined by the symmetry of locally Minkowski spacetime is also different from the Euclidean length. So does the unit circle and its related quantity-$\pi$. Here, we denote the Finslerian $\pi$ by $\pi_F$. `$\wedge$' is the `wedge product'. For more details please refer to the book \citep{Book by Chern}.}.
The solution of (\ref{static field eq2},\ref{static field eq3}) is given as
\begin{equation}\label{static field eq4}
h_{00}=-\frac{2GM}{R}\eta_{00},~~ h_{nn}=\frac{2GM}{R}\eta_{nn},
\end{equation}
where $R^2\equiv\eta_{ij}x^i x^j$.
In Newton's limit, the geodesic equation (\ref{geodesic}) reduces to
\begin{eqnarray}\label{geodesic compo1}
\frac{d^2x^0}{d\tau^2}-\frac{\eta^{0i}}{2}\frac{\partial h_{00}}{\partial x^i}\frac{dx^0}{d\tau}\frac{dx^0}{d\tau}=0,\\
\label{geodesic compo2}
\frac{d^2x^i}{d\tau^2}-\frac{\eta^{ij}}{2}\frac{\partial h_{00}}{\partial x^j}\frac{dx^0}{d\tau}\frac{dx^0}{d\tau}=0.
\end{eqnarray}
The equation (\ref{geodesic compo1}) implies that $\frac{dx^0}{d\tau}$ is a function of $h_{00}$. Since $|h_{00}|\ll1$, $\frac{dx^0}{d\tau}$ could be treated as a constant in equation (\ref{geodesic compo2}). Then, we find from (\ref{geodesic compo2}) that
\begin{equation}\label{law of gravity}
\frac{d^2x^i}{{dx^0}^2}=-\frac{GM}{R^2}\frac{x^i}{R},
\end{equation}
where ${dx^0}^2=\eta_{00}dx^0 dx^0$. The formula (\ref{law of gravity}) implies that the law of gravity in Finsler spacetime is similar to that in Newton's case. The difference is that the spatial distance is now Finslerian. It is what we expect from Finslerian gravity, because the length difference is one of the major attributes of Finsler geometry as compared to the Riemannian geometry.

\subsection{Finslerian MOND}
In \citep{Finsler MOND}, we have shown that Finsler gravity reduces to MOND, if the spacial part of the locally Minkowski metric of galaxies is of the form
\begin{equation}\label{length gala1}
\eta_{ij}=\delta_{ij}\left(1-\left(\frac{GMa_0{y^0}^4}{(\delta_{mn}y^m y^n)^2}\right)^2\right),
\end{equation}
where $a_0=1.2\times10^{-10}{\rm m/s^2}$ is the constant of MOND. In Finsler spacetime, the speed of particle is given as  $v^i\equiv \frac{dx^i}{dx^0}=\frac{y^i}{y^0}$. The radial coordinate in the locally Minkowski space-time of galaxies (\ref{length gala1}) can be written as
\begin{equation}\label{length gala2}
R\equiv \sqrt{\eta_{ij}x^i x^j}=r\sqrt{1-\left(\frac{GMa_0}{v^4}\right)^2}\equiv rf(v),
\end{equation}
where $r^2=\delta_{ij}x^i x^j$ and $v^2\equiv\delta_{ij}v^i v^j$. Substituting (\ref{length gala2}) back into (\ref{law of gravity}), we obtain the result of MOND
\begin{equation}\label{MOND}
\frac{GM}{r^2}=\frac{v^2}{r}\mu\left(\frac{v^2}{ra_0}\right),
\end{equation}
where $\mu(x)=x/\sqrt{x^2+1}$ is the interpolating function in MOND.

In this paper, we try to consider multipole effects of Finslerian MOND, and use them to explain the observed $\kappa$-map of Bullet Cluster. The Finslerian radial coordinate has the form $R=rf(v)$. And without losing any generality in our discussion of the motion of particle in Finsler spacetime, we set $\eta_{00}$ to 1. Then, we obtain the Finsler structure in the post-Newtonian approximation from (\ref{static field eq4})
\begin{equation}\label{length deflection}
F^2 d\tau^2=\left(1-\frac{2GM}{R}\right)d\tau^2-\left(1+\frac{2GM}{R}\right)dR^2-R^2(d\theta^2+\sin^2\theta d\phi^2).
\end{equation}

To first order in $h$, the geodesic spray coefficients are
\begin{equation}\label{geodesic spray appro}
G^\mu=\frac{1}{4}\eta^{\mu\nu}\left(2\frac{\partial h_{\alpha\nu}}{\partial x^\lambda}y^\alpha y^\lambda-\frac{\partial h_{\alpha\beta}}{\partial x^\nu}y^\alpha y^\beta\right).
\end{equation}
Then, given (\ref{geodesic spray appro}), one can solve the geodesic equation (\ref{geodesic}). And by making use of the stipulation $F=0$ in (\ref{length deflection}) for photons, one could obtain the formula of gravitational deflection of light in Finsler spacetime. We skip the conventional calculations here. In fact, the Finslerian line element (\ref{length deflection}) is simply the Schwarzschild's one except for a rescaling of the Euclidean radial coordinate $r$. It is also true for the geodesic equation. Thus, the deflection angle in Finsler gravity is a rescaling of Einstein's one. It is of the form
\begin{equation}\label{Finsler deflect}
\alpha_F=\frac{4GM}{R_m}, ~~~~R_m = r_mf(v_m),
\end{equation}
where $v_m$ is the fiber coordinate for corresponding $r_m$. $R_m$ is the closest distance of the light path to the gravitational source, where $r_m$ is that in a Euclidean space.

\section{Convergence $\kappa$ in Finsler Gravity}
The ICM gas of Bullet Cluster can be well described by the King $\beta$-model \citep{Cavaliere}. Its mass density distribution is given as
\begin{equation}\label{king beta}
\rho(r)=\rho_0\left(1+\frac{r^2}{r^2_c}\right)^{-3\beta/2}.
\end{equation}
The surface mass density is given by integrating $\rho(r)$ of equation (\ref{king beta}) along the line of sight
\begin{equation}
\Sigma(x_1,x_2)=\int^{x_3^{\rm{out}}}_{-x_3^{\rm{out}}} \rho(x_1,x_2,x_3) dx_3~.
\end{equation}
In the limit $x_3^{\rm{out}}\gg r_c$, the surface mass density is of the form (detailed discussions can be found in \citep{Moffat})
\begin{equation}\label{Sigma map}
\Sigma(\xi)=\Sigma_0\left(1+\frac{\xi^2}{r_c^2}\right)^{-(3\beta-1)/2},
\end{equation}
where $\xi^2\equiv x^2_1+x^2_2$ is defined in the lens plane and $\Sigma_0=\sqrt{\pi}\rho_0 r_c\Gamma(\frac{3\beta-1}{2})/\Gamma(\frac{3\beta}{2})$.
Moffat {\it et al}. For the observed ICM gas profile of the main cluster, the best-fit parameters of the King $\beta$-model (\ref{king beta}) are given as \citep{Moffat}
\begin{equation}\label{king para}
\beta=0.803\pm0.013,~~~r_c=278.0\pm6.8~{\rm kpc},~~~\rho_0=3.34\times10^5~ {\rm M_\odot/kpc^3}.
\end{equation}

In observations, the convergence $\kappa_G$ measures the ratio of observed surface density to the critical surface density \citep{Peacock}
\begin{equation}
\kappa_G=4\pi G\int \frac{D_LD_{LS}}{D_S} \rho(x_1,x_2,x_3)dx_3\equiv\frac{\Sigma(x_1,x_2)}{\Sigma_c},
\end{equation}
where $\Sigma_c=\frac{1}{4\pi G}\frac{D_S}{D_LD_{LS}}$, $D_S$ is the distance between the source galaxy and the observer, $D_L$ is the distance between the lens (Bullet Cluster) and the observer, and $D_{LS}$ is the distance between the source galaxy and the lens. For the Bullet Cluster, the critical surface density $\Sigma_c$ takes a value of $3.1\times10^9 {\rm M_\odot/kpc^2}$ \citep{Clowe,Moffat}.

In general relativity, the convergence $\kappa$ maps the gravitational lensing effect of a given gravitational source.
It can be expressed in terms of the deflection angle as
\begin{equation}
\kappa=\frac{1}{2}\frac{D_{LS}D_L}{D_S}\nabla_\xi\alpha\ ,
\label{kappa in deflection angle}
\end{equation}
where $\xi=\sqrt{x_1^2+x_2^2}$.
The deflection angle in Finsler spacetime was given as (\ref{Finsler deflect}). Substituting $\alpha_F$ into (\ref{kappa in deflection angle}), we obtained
\begin{equation}
\kappa_F=\frac{1}{2}\frac{D_{LS}D_L}{D_S}\left[\frac{1}{f(v)}\nabla_\xi \alpha_G + \alpha_G \nabla_\xi \frac{1}{f(v)}\right]=\frac{1}{f(v)}\kappa_G + \frac{1}{2}\frac{D_{LS}D_L}{D_S}\alpha_G \nabla_\xi \frac{1}{f(v)}\ ,
\label{full Finslerian kappa}
\end{equation}
where
\begin{equation}
\kappa_G \equiv \frac{1}{2}\frac{D_{LS}D_L}{D_S}\nabla_\xi \alpha_G\, .
\end{equation}
The first term is simply a rescaling of $\kappa$ given by general relativity. The second term depends on the specific form of $f$. It does not retain the linearity and the superposition principle of the point mass potential on mass $m$. It can also be neglected in the weak field approximation.
We will demonstrate the second point in APPENDIX and show that for the two cases of $f$ (see the next section) that were investigated in this paper, the second term is a few orders smaller than the first term and thus can be neglected. The $\kappa_F$ can be approximately given by
\begin{equation}
\kappa_F\simeq\frac{1}{f(v)}\kappa_G\, .
\label{Finsler kappa}
\end{equation}

Here, we summarize the logic steps to deduce the convergence $\kappa_F$ in Finsler gravity. First, we extended Pirani's argument of equation of motion to the case of Finsler geometry to get $Ric=0$, which can be derived from an action integral on the unit tangent bundle \citep{Pfeifer1}. Second, in post-Newtonian approximation, we obtained gravitational field equation in Finsler gravity (\ref{static field eq1}). Third, in the Newtonian limit to first order of $GM/R$, we obtained the Finsler line element (\ref{length deflection}). It is simply the Schwarzschild's metric except for the rescaling coefficient $f(v)$ of the Euclidean radial coordinate. Then, we obtained the deflection angle (\ref{Finsler deflect}) in Finsler gravity. Fourth, given the relation (\ref{kappa in deflection angle}) between the convergence $\kappa$ and the deflection angle $\alpha$, we obtained the Finslerian convergence $\kappa_F$ as (\ref{Finsler kappa}). Up till now, our formulae in Finsler gravity haven been presented on the tangent bundle. However, the physics of the astronomical observations lie in four-dimensional spacetime. We need a projection that translates the formulae on the tangent bundle into the ones on the manifold. Such a projection stems from the solution of geodesic equation. The geodesic equation (\ref{geodesic}) gives the relation $y\equiv\frac{dx}{d\tau}=y(x)$. It implies that $f(v)$ could be written as a function of $x$ by the relation $y(x)$. Finally, after doing these steps, we obtain the Finslerian convergence
\begin{equation}\label{Finsler kappa1}
\kappa_F\simeq\frac{1}{g(x)}\kappa_G,
\end{equation}
where $g(x)\equiv f(y(x))$. Given the surface mass density profile (\ref{Sigma map}), we could obtain the numerical results of convergence $\kappa$-map from the equation (\ref{Finsler kappa1}).

\section{Numerical Analysis}
\subsection{The Convergence $\kappa$-Map}
\begin{figure}
\begin{center}
\includegraphics[scale=0.7]{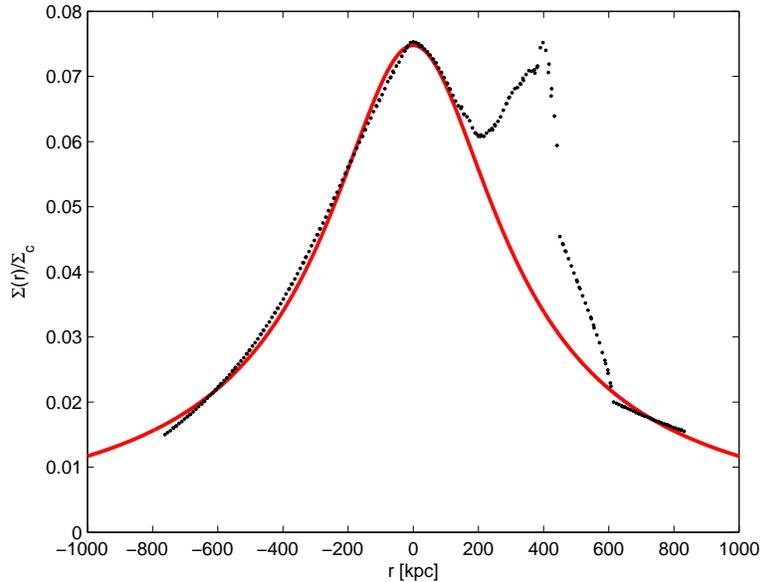}
\caption{The scaled $\Sigma$-map from X-ray imaging observations of the Bullet Cluster 1E0657-558, November 15, 2006 data
release \citep{Bullet2,Bullet}. The peak of the main cluster is taken to be the  referential center of the system, i.e. $r=0$ . The peak of the subcluster locates at $\xi\simeq398$ kpc. A cross-section of the observed $\Sigma$-map, on a straight-line connecting the peak of the main cluster to that of the subcluster, is shown in solid black dots. The best-fit King $\beta$-model for the surface mass density is shown in solid red. Negative radii have no particular meanings but only denote the left-hand-side of the map relative to the origin $r=0$. They have the same physical interpretation as the positive-half radii after being placed an absolute sign `$|~~|$'.}
\label{fig1}
\end{center}
\end{figure}

The surface mass density profile (\ref{Sigma map}) with best-fit parameters (\ref{king para}) are shown in Figure \ref{fig1}. The main X-ray cluster is set at $\xi=0~{\rm kpc}$ and the subcluster (the peak of which lies at $\xi\sim400~{\rm kpc}$) is neglected in doing the best-fit with the King $\beta$-model. The surface mass density (\ref{Sigma map}) for the main cluster includes most of the ICM gas. It implies that the ICM gas profile of the Bullet Cluster is in approximate spherical symmetry. We will use the surface mass density (\ref{Sigma map}) to calculate the convergence $\kappa_F$.

In this paper, our motivation is to construct a MOND-like theory in Finsler gravity, and use it to explain the observations of the Bullet Cluster. As mentioned in the introduction, a modified gravity theory is taken as a theory of MOND so long as it reduces to Newton's gravity while the MONDian constant $a_0\rightarrow0$ and the Tully-Fisher relation holds for deep-MOND limit, $a_0\rightarrow\infty$.

\begin{figure}
\begin{center}
\includegraphics[scale=0.7]{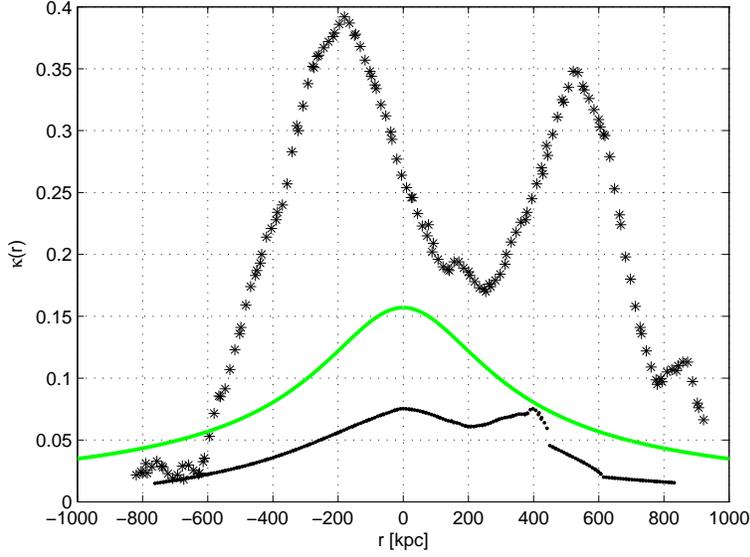}
\caption{The $\kappa$-map reconstructed from the strong and weak gravitational lensing survey of the Bullet Cluster 1E0657-558, November 15, 2006 data release \citep{Bullet2,Bullet}. The solid black dots denote the cross-section of the scaled $\Sigma$-map from the X-ray imaging observations that presented in Figure \ref{fig1}. A section of the reconstructed $\kappa$-map, on a straight-line connecting the peak of the main cluster to that of the subcluster, is shown in black stars. The peak of the main cluster locates at $\xi \simeq -180$ kpc and that of the subcluster locates at $\xi\simeq 522$ kpc. The $\xi=0$ point is chosen to be the same with that of the $\Sigma$-map in Figure \ref{fig1}. The convergence $\kappa_F$ predicted by the Finslerian MOND is shown in solid green, for which the Finslerian factor $g_M(r)$ is of the form (\ref{monopole g}). The interpretation of negative radii is the same as that in Figure \ref{fig1}.}
\label{fig2}
\end{center}
\end{figure}

First, we propose a Finslerian MOND with spherical symmetry. The geodesic equation gives an approximate relation between the velocity and the modified gravitational potential \citep{Finsler MOND}
\begin{equation}\label{vR relation}
v^2=\frac{GM}{R}=\frac{GM}{rf(v)}.
\end{equation}
If
\begin{equation}\label{monopole g}
g_M(r)\equiv f(v(r))=\left(1+\sqrt{\frac{a_0 r^2}{GM}}~\right)^{-1},
\end{equation}
we find from (\ref{vR relation}) that
\begin{equation}
v^2=\frac{GM}{r}+\sqrt{GMa_0}\, .
\end{equation}
It is a MOND theory with spherical symmetry. It should be noticed that the three-dimensional radial distance $r$ equals the two-dimensional radial distance $\xi$ if one deals with the physics in the lens plane. So, in this section, we use $r$ to represent the radial distance on the lens plane. Substituting (\ref{monopole g}) into (\ref{Finsler kappa1}), we obtain the convergence $\kappa$-map given by MOND theory with spherical symmetry. The result is shown in Figure \ref{fig2}. In Figure \ref{fig2}, one can find that a MOND theory with spherical symmetry cannot account for the reconstructed convergence $\kappa$-map of Bullet Cluster. The convergence $\kappa$-map of Bullet Cluster shows that the distribution of gravitational force is anisotropic. To describe the anisotropic force, we should introduce multipole fields. The dipole contribution vanishes if one takes the center of ICM gas as the coordinate origin. In fact, Milgrom gives a quasi-linear formulation of MOND (QUMOND)\citep{QUMOND}, which involves the quadrupole contribution.

\begin{figure}
\begin{center}
\includegraphics[scale=0.55]{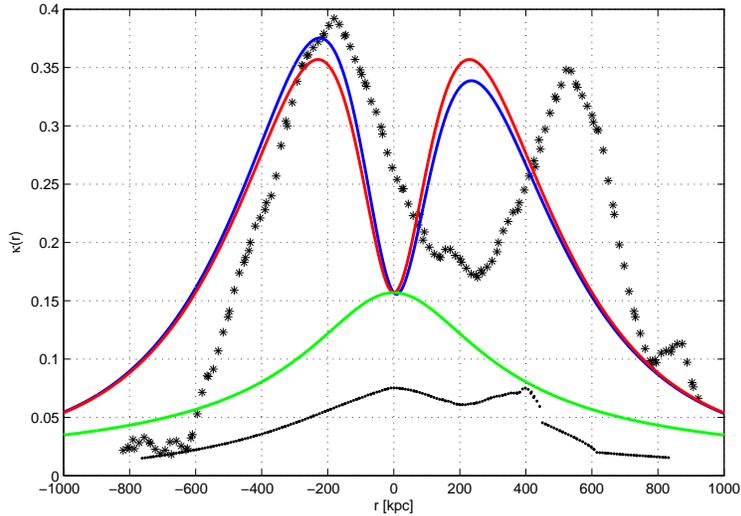}
\caption{A cross-section on a straight line connecting the two peaks of the reconstructed $\kappa$-map reconstructed from the strong and weak gravitational lensing survey of Bullet Cluster 1E0657-558, November 15, 2006 data release \citep{Bullet2,Bullet}. The solid black dots denote the cross-section of the scaled $\Sigma$-map from the X-ray imaging observations that presented in Figure \ref{fig1}. Quadrupole contribution is considered in the Finslerian MOND in the prediction of convergence $\kappa_F$, which is shown in solid red. The result that taking the subcluster's contribution of ICM gas into account as well as the quadrupole contribution is shown in solid blue, for which the Finslerian factor $g_{QD}(r)$ is of the form (\ref{quadrupole plus dipole g}). The interpretation of negative radii is the same as that in Figure \ref{fig1}, in addition that now the negative-half correspond to $\theta=0$ while the positive-half require that $\theta=\pi$.}
\label{fig3}
\end{center}
\end{figure}

Here comes our second step. We take the quadrupole effect into consideration in Finslerian MOND in a way that the quadrupole contribution appears even at large scales. The Finslerian parameter $g_Q(r,\theta)$ now takes the form
\begin{equation}\label{quadrupole g}
g^{-1}_Q(r,\theta)=1+\sqrt{\frac{a_0 r^2}{GM}}\left(1+\frac{GMa_0}{b^4}\cos^2\theta\exp(-r/c)\right),
\end{equation}
where the parameters $b=458$ km/s and $c=220$ kpc.
In order to keep the Tully-Fisher relation, an exponential term $\exp(-r/c)$ is needed in (\ref{quadrupole g}).
Substituting (\ref{quadrupole g}) into (\ref{Finsler kappa1}), we obtain the convergence $\kappa$-map giving by MOND theory with monopole contribution plus quadrupole contribution. The result is shown in Figure \ref{fig3}. The monopole contribution plus quadrupole contribution can account for the main feature of the convergence $\kappa$-map of Bullet Cluster, except for the asymmetry between the convergence of the main cluster and the subcluster. Until now, we only consider the effect of the spherical part (i.e. the main cluster of the $\Sigma$-map) of ICM gas. The dipole contribution vanishes as we take the coordinate origin to be the center of ICM gas.

\begin{figure}
\begin{center}
\includegraphics[scale=0.9]{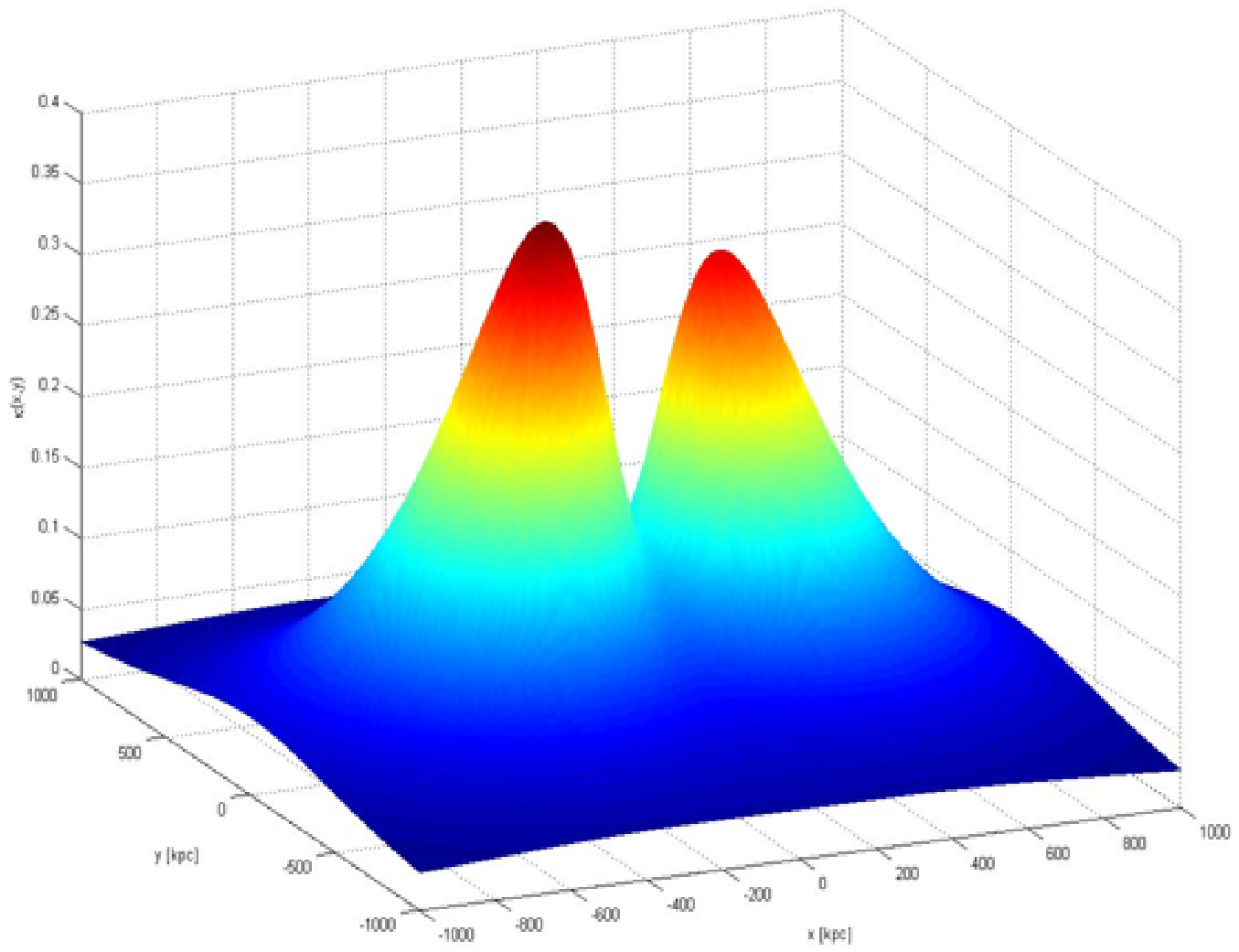}
\caption{ The 3D figure of convergence $\kappa_F$ given by the Finslerian MOND, where the Finslerian parameter $g_{QD}(r)$ is of the form (\ref{quadrupole plus dipole g}). The interpretation of negative radii is the same as that in Figure \ref{fig1}.}
\label{fig4}
\end{center}
\end{figure}

\begin{figure}
\begin{center}
\includegraphics[scale=0.9]{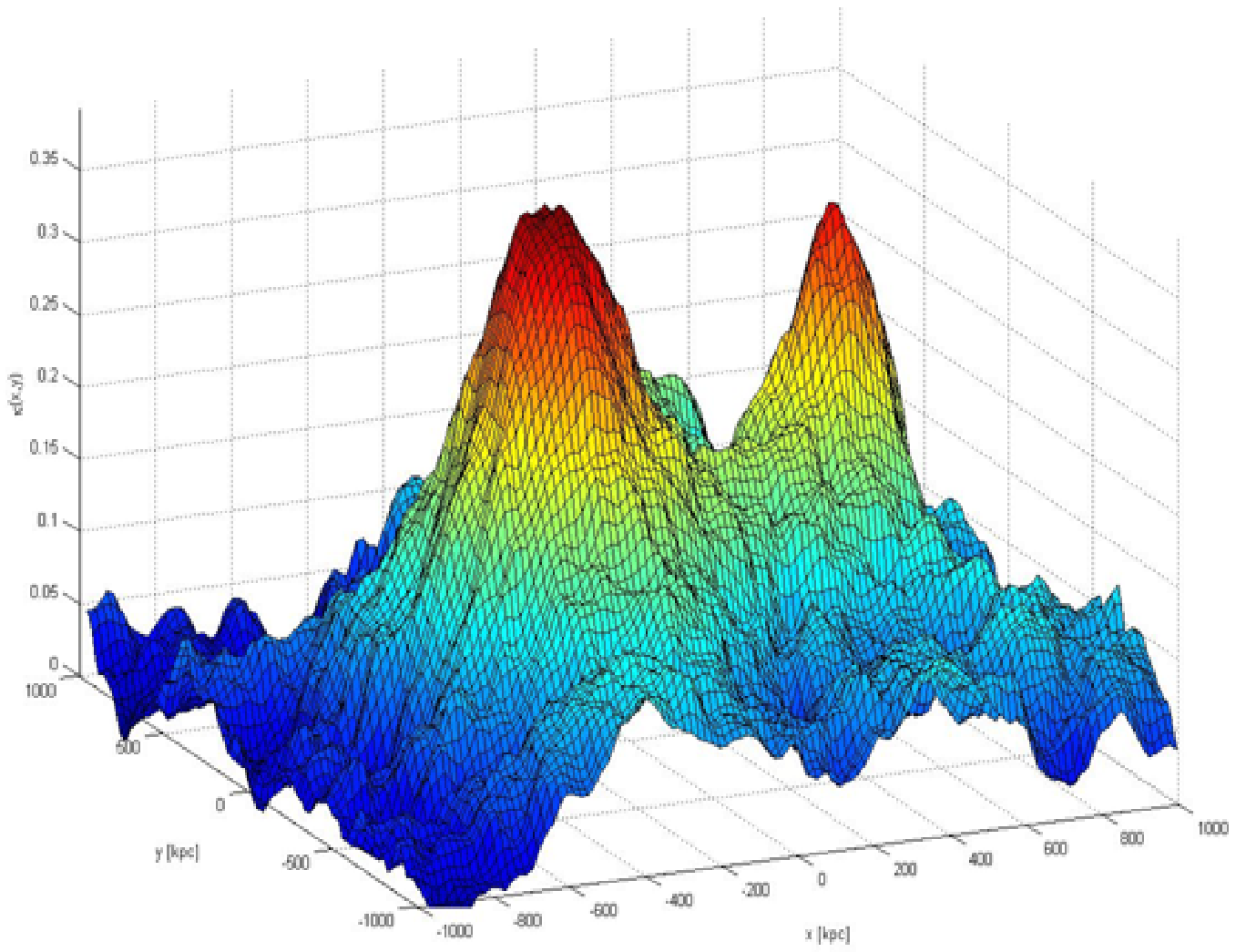}
\caption{The 3D $\kappa$-map reconstructed from the strong and weak gravitational lensing survey of the Bullet Cluster 1E0657-558, November 15, 2006 data release \citep{Bullet2,Bullet}. The interpretation of negative radii is the same as that in Figure \ref{fig1}.}
\label{fig5}
\end{center}
\end{figure}

Here comes our final step, we consider the subcluster of the $\Sigma$-map and regard it as a perturbation. Equivalently, it could be regarded as a dipole contribution. Then, The Finslerian parameter $g(r,\theta)$ is of the form
\begin{equation}\label{quadrupole plus dipole g}
g^{-1}_{QD}(r,\theta)=1+\sqrt{\frac{a_0 r^2}{GM}}\left(1+\frac{\sqrt{GMa_0}}{a^2}\cos\theta\exp(-r/c)+\frac{GMa_0}{b^4}\cos^2\theta\exp(-r/c)\right),
\end{equation}
where the parameter $a=2b\simeq916$ km/s. In formula (\ref{quadrupole plus dipole g}), the dipole term $\frac{\sqrt{GMa_0}}{a^2}\simeq 1$ for $r\simeq780 ~{\rm kpc}$. It means that the dipole term in (\ref{quadrupole plus dipole g}) becomes dominant at $r\simeq 780$ kpc. Distance at this far almost reaches the boundary of the Bullet Cluster system. And it is suppressed by the exponential term $\exp(-r/c)$. Thus, it is justified to regard the dipole term in (\ref{quadrupole plus dipole g}) as a perturbation.
Substituting (\ref{quadrupole plus dipole g}) into (\ref{Finsler kappa1}), we obtain the convergence $\kappa$ which is predicted by the Finslerian MOND theory with the contribution of a monopole, a quadrupole and that of a dipole perturbation. The result is also shown in Figure \ref{fig3}. One can see the asymmetry between the convergence $\kappa$ peak of the main cluster and the subcluster, with the center of convergence $\kappa_F$ for the system lying at a few {\rm kpc}s away from the origin due to the dipole effect. The 3D figure is shown in Figure \ref{fig4} and the observational data of the convergence $\kappa$ of Bullet Cluster is presented in Figure \ref{fig5} for comparison.

In the last section, we concluded that for our discussions, the equation (\ref{full Finslerian kappa}) can be well approximated by equation (\ref{Finsler kappa}). Given the (\ref{monopole g}) and (\ref{quadrupole plus dipole g}), we check this point carefully in APPENDIX.

\subsection{The Isothermal Spherical Mass Profile}
At last, we give a discussion about the isothermal temperature of the ICM gas profile of the main cluster in Finsler gravity.
The ICM gas profile of the main cluster can be regarded as a spherical and isotropic system. If it is in hydrostatic equilibrium, it satisfies the collisionless Boltzmann equation
\begin{equation}\label{Boltzmann eq}
a(r)\equiv-\frac{d\Phi}{dr}=\frac{1}{\rho}\frac{d\rho\sigma_r^2}{dr},
\end{equation}
where $\Phi$ is the gravitational potential and $\sigma_r$ is the velocity dispersion. Assuming an isothermal gas profile, $\sigma_r$ is related to the isothermal temperature $T$ of the ICM gas as
\begin{equation}
\sigma_r^2=\frac{kT}{\mu m_p}\ ,
\label{velocity dispersion}
\end{equation}
where $k$ is the Boltzmann constant, $\mu\approx 0.609$ is the mean atomic weight and $m_p$ is the proton mass. Substituting both formula (\ref{velocity dispersion}) and the density distribution (\ref{king beta}) of King $\beta$-model into (\ref{Boltzmann eq}), we obtain that
\begin{equation}\label{tempareture a}
a(r)=-\frac{3\beta T}{\mu m_p}\left( \frac{r^2}{r^2+r^2_c}\right).
\end{equation}

\begin{figure}
\begin{center}
\includegraphics[scale=0.7]{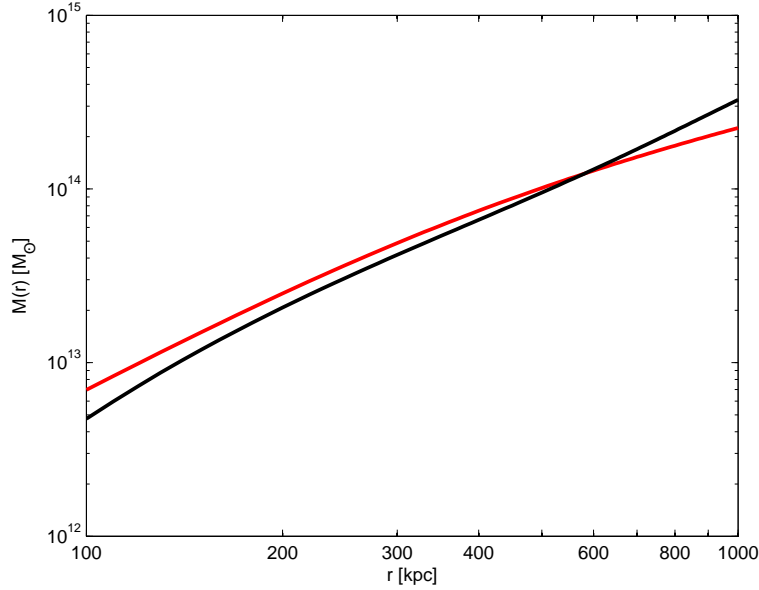}
\caption{The mass profile given by the best-fit King $\beta$-model of the maincluster is shown in solid red. The mass profile derived from Finslerian MOND with quadrupole effect is shown in solid black. The results are presented on a logarithmic scale for both the $M$- and $r$-axis.}
\label{fig6}
\end{center}
\end{figure}

Markevitch {\it et al}. \citep{Markevitch} have presented the experimental value of the isothermal temperature of the main
cluster $T=14.8^{+1.7}_{-2.0}$ keV with $4.5\%$ error. By making use of (\ref{tempareture a}), one can find that the Newton's gravitational force $a_N=-GM/r$ cannot provide enough force to maintain the hydrostatic equilibrium. In Finsler gravity, the gravitational acceleration law is of the form
\begin{equation}\label{Finsler acc}
a_F=-\frac{GM}{R^2}=-\frac{GM}{(rg(r))^2}.
\end{equation}
We neglect the dipole perturbation in our study of the isothermal temperature of the main cluster and only consider the quadrupole contribution in Finslerian MOND. Even this, the gravitational system is no more isotropic. Nevertheless, we could take the average of the radial force by integrating $g(r,\theta)$ of (\ref{quadrupole g}) over $\theta$
\begin{equation}\label{average g}
\bar{g}(r)^{-1}=\frac{1}{2\pi}\int^{2\pi}_0 g(r,\theta)^{-1}d\theta=1+\sqrt{\frac{a_0 r^2}{GM}}\left(1+\frac{GMa_0}{2b^4}\exp(-r/c)\right),
\end{equation}
and use it to qualitatively study the hydrostatic equilibrium of an isotropic system.
Substituting the $\bar{g}(r)$ of equation (\ref{average g}) into (\ref{Finsler acc}), we obtain that
\begin{equation}\label{Finsler acc1}
a_F=-\frac{GM}{r^2}\left(1+\sqrt{\frac{a_0 r^2}{GM}}\left(1+\frac{GMa_0}{2b^4}\exp(-r/c)\right)\right)^2\, .
\end{equation}
Here, we take the temperature in formula (\ref{tempareture a}) to be $14.8$ keV, which is the experimental mean value given by Markevitch {\it et al}. \citep{Markevitch}. Then, by identifying the equation (\ref{tempareture a}) with (\ref{Finsler acc1}), we obtain the mass profile of the main cluster of ICM gas in Finslerian MOND. In Figure \ref{fig6}, we compare it with the result of the best-fit King $\beta$-model. It is shown that the two mass profiles have the same order. It means that the Finslerian MOND with quadrupole effect agree with the observations \citep{Markevitch}.

\section{Conclusions and Discussions}
In this paper, we try to setup a Finslerian MOND, a generalization of MOND in Finsler spacetime. We extended Pirani's argument to get the stipulation $Ric=0$, from which we obtained the gravitational vacuum field equation in Finsler spacetime. Considering the correspondence with the post-Newtonian limit of general relativity, we got the explicit form of the Finslerian line element. It was simply the Schwarzschild's metric except for the Finslerian rescaling coefficient $f(v)$ of the radial coordinate $r$. Given that $f(v)=\sqrt{1-(GMa_0/v^4)}$, we recovered the famous MOND in a Finslerian framework. By introducing a quadrupole and a dipole perturbation term into the Finslerian MOND, we calculated the convergence $\kappa$ in gravitational lensing astrophysics. A qualitative-level numerical analysis showed that our prediction is in agreement with the observed $\kappa$-map of Bullet Cluster 1E0657-558. Given the observed value 14.8 keV of the isothermal temperature of the main cluster in our model, the predicted mass density profile of the main cluster is the same order as that given by the best-fit King $\beta$-model.

However, one should notice that the factor $f(v)$ (i.e. $g(r)$) is determined by the local spacetime symmetry, which cannot be deduced from the gravity theory. It is not the fruit but a prior stipulation of the theory. The logic is: given a specific $f(v)$, we then proceed to calculate the convergence $\kappa$-map and the temperature of the main cluster. The coefficient $f(v)$ in our model comes directly from the flat Finsler spacetime $\eta_{\mu\nu}(y)$ (\ref{expand metric}). In fact, while the Euclidean radial distance $r\rightarrow\infty$, the Finslerian length element (\ref{length deflection}) reduces to
\begin{equation}
F^2 d\tau^2=d\tau^2-dR^2-R^2(d\theta^2+\sin^2\theta d\phi^2).
\end{equation}
It is simply the line element of flat Finsler spacetime $\eta_{\mu\nu}(y)$. The coefficient $f(v)$ could be arbitrary in principle, since we suppose the metric is close to the flat Finsler spacetime $\eta_{\mu\nu}(y)$.
Most of the galaxies could be regarded as a spherical system and described by a central modified gravitational potential. Therefore, to describe it in Finsler gravity, the Finsler parameter $f(v)$ should be spherical. It means that the flat Finsler spacetime $\eta_{\mu\nu}(y)$ is not universal in cosmology. At present, the specific form of flat Finsler spacetime $\eta_{\mu\nu}(y)$ could be regard as an axiom in our theory of Finsler gravity. There is no physical equation or principle to constrain the form of it. \textbf{Professor Shen's} description of Finsler geometry (private conversation) may help us in understanding what is a flat Finsler spacetime $\eta_{\mu\nu}(y)$ --- ``Riemann geometry is `a white egg', for the tangent manifold at each point on the Riemannian manifold is isometric to a Minkowski spacetime. However, Finsler geometry is `a colorful egg', for the tangent manifolds at different points of the Finsler manifold are not isometric to each other in general.'' In physics, it implies that our nature does not always prefer an isotropic gravitational force. It is also ``colorful'', as we have seen in case of Bullet Cluster 1E0657-558.

\section*{Appendix}
We will show that for the two cases of $f$ (or $g$) that were investigated in this paper, the second term in Eq.(46) are both a few orders smaller than the first term and thus can be neglected.

Given that $g_M(r)\equiv f(v(r))=\left(1+\sqrt{\frac{a_0 r^2}{GM}}~\right)^{-1}$ for MOND, we get $\kappa_F=\frac{1}{f(v(r))}\kappa_G + \frac{1}{2}\frac{D_{LS}D_L}{D_S} \sqrt{\frac{a_0}{GM}}\alpha_G$. To get this result, we have replaced $r$ with $\xi=\sqrt{x_1^2+x_2^2}$. For a rough estimate of the magnitude of the second term, we write $\alpha_G$ as $\alpha_G=\frac{4GM}{c^2 \xi}$, where $c\simeq3\times10^{8}~\textmd{m/s}=9.71\times10^{-12}~\textmd{kpc/s}$ is the speed of light in vacuum. For the Bullet Cluster system, we have a total mass of $M\simeq 10^{14}M_\odot$ and a distance range of $0\leq \xi \leq 1000$ kpc. The constant for MOND is $a_0=1.2\times 10^{-8}~\textmd{cm/s}^2 \simeq3.84\times 10^{-30}~\textmd{kpc/s}^2$. A simple arithmetic exercise shows that the magnitude of the second term in the expression of $\kappa_F$ (i.e. the term $\frac{1}{2}\frac{D_{LS}D_L}{D_S} \sqrt{\frac{a_0}{GM}}\alpha_G$) is at about $10^{-5}$, which can be neglected comparing to the first term $\frac{1}{f(v(r))}\kappa_G$, of which value ranges from $0.1\sim 0.4$.

The calculation in the last paragraph was carried assuming that $M$ is a constant. If mass $M$ is taken as a function of $\xi$, i.e. $M=M(\xi)$, $\kappa_F$ takes a form $\kappa_F=\frac{1}{f(v(r))}\kappa_G + \frac{1}{2}\frac{D_{LS}D_L}{D_S} \sqrt{\frac{a_0}{GM}}\alpha_G + \left(\frac{-D_{LS}D_L}{4D_S }\right)\alpha_G\sqrt{\frac{a_0\xi^2}{GM}}\cdot \frac{dM}{d\xi}/M$. Using $\alpha_G=\frac{4GM}{c^2 \xi}$, the third term on the right hand side of the above identity can be reduced into $\left(\frac{-D_{LS}D_L}{D_S}\right)\frac{\sqrt{GMa_0}}{c^2}\cdot \frac{dM}{d\xi}/M$. The term $\frac{dM}{d\xi}/M$ has an order of $10^{-3}$ kpc$^{-1}$ and $\sqrt{GMa_0}\simeq 10^{-27}$ $\textmd{(kpc/s)}^2$, giving that the whole third term $\left(\frac{-D_{LS}D_L}{D_S}\right)\frac{\sqrt{GMa_0}}{c^2}\cdot \frac{dM}{d\xi}/M \simeq 10^{-6}$, which is negligible comparing to the first two terms in the expression of $\kappa_F$. Therefore, for $g_M(r)\equiv f(v(r))=\left(1+\sqrt{\frac{a_0 r^2}{GM}}~\right)^{-1}$, the convergence $\kappa_F$ is given as $\kappa_F\simeq\frac{1}{f(v(r))}\kappa_G$.

\begin{figure}
\begin{center}
\subfigure[~\textsf{$g^{-1}_{QD}(\xi,\theta)$ vs. $\xi$}] {\label{fig:a}\scalebox{0.6}{\includegraphics{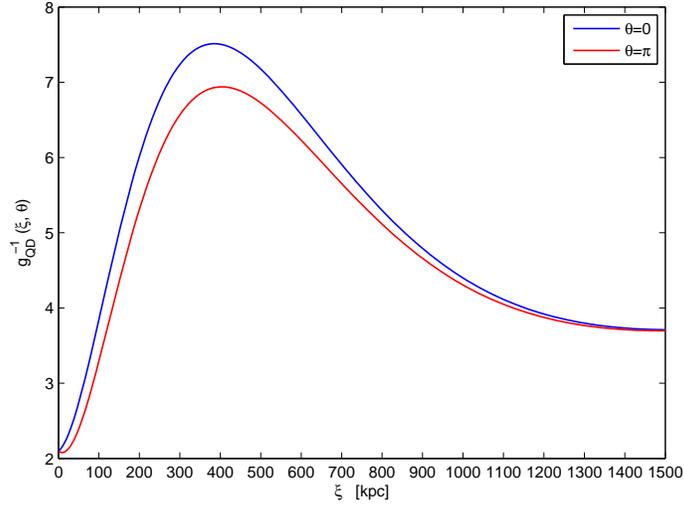}}}
\subfigure[~\textsf{$\nabla_\xi g^{-1}_{QD}(\xi,\theta)$ vs. $\xi$}] {\label{fig:b}\scalebox{0.6}{\includegraphics{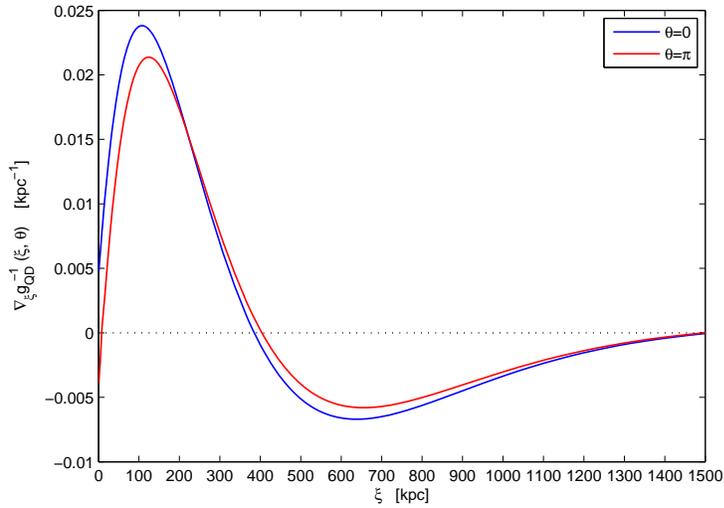}}}
\caption{Plot of $g^{-1}_{QD}(\xi,\theta)$ and $\nabla_\xi g^{-1}_{QD}(\xi,\theta)$ vs. distance $\xi$ in kpc. The blue solid line represents that for $\theta=0$ and the red one for $\theta=\pi $. One can see that $\nabla_\xi g^{-1}_{QD}(\xi,\theta)\simeq 10^{-2}$ kpc$^{-1}$.}
\label{fr}
\end{center}
\end{figure}

This conclusion still holds true for the anisotropic Finslerian MOND model we presented. Given that $g_{QD}(r,\theta)^{-1}=1+\sqrt{\frac{a_0 r^2}{GM}}\left(1+\frac{\sqrt{GMa_0}}{a^2}\cos\theta\exp(-r/c)+\frac{GMa_0}{b^4}\cos^2\theta\exp(-r/c)\right)$, we obtain the corresponding $\kappa$ as $\kappa_{QD}=\frac{1}{g_{QD}(\xi,\theta)}\kappa_G + \frac{1}{2}\frac{D_{LS}D_L}{D_S}\alpha_G \nabla_\xi \frac{1}{g_{QD}(\xi,\theta)}$. We plot $g_{QD}^{-1}(\xi,\theta)$ and $\nabla_\xi \frac{1}{g_{QD}(\xi,\theta)}$ as functions of $\xi$ respectively in Figure \ref{fr}. One can see that $\nabla_\xi \frac{1}{g_{QD}(\xi,\theta)}\simeq 10^{-2}$ kpc$^{-1}$. Again we can check that for the given parameters $a=2b\simeq916$ km/s and $c=220$ kpc, the term $\frac{1}{2}\frac{D_{LS}D_L}{D_S}\alpha_G \nabla_\xi \frac{1}{g_{QD}(\xi,\theta)}\simeq10^{-5}$ is also too small comparing to the first term $\frac{1}{f(v(r,\theta))}\kappa_G$. Thus it is justified to be neglected in the calculation of $\kappa_F$.

Therefore, considering the above discussions, the convergence $\kappa_F$ in our Finsler gravity model is given as $\kappa_F\simeq\frac{1}{f(v(r,\theta))}\kappa_G$.

\section*{Acknowledgments}

We would like to thank S. Wang and Y.-G. Jiang for useful discussions. The work was supported by the NSF of China under Grant No. 11075166 and No. 11147176.

\bsp

\label{lastpage}

\end{document}